\documentclass[aps,twocolumn,pra,showpacs,amsmath,nofootinbib]{revtex4}
\usepackage{graphicx}
\usepackage{dcolumn}
\usepackage{bm}
\usepackage{epstopdf}
\usepackage{multirow}
\newcommand{\beq}{\begin{equation}}
\newcommand{\eeq}{\end{equation}}
\newcommand{\bea}{\begin{eqnarray}}
\newcommand{\eea}{\end{eqnarray}}

\begin{document}
\author{S. Kamerdzhiev}
\affiliation{National Research Centre "Kurchatov Institute", Moscow, Russia}
\author{O. Achakovskiy\footnote{oachakovskiy@ippe.ru}}
\affiliation{Institute for Physics and Power Engineering, 249033 Obninsk, Russia}
\author{A. Avdeenkov}
\affiliation{Institute for Physics and Power Engineering, 249033 Obninsk, Russia}
\author{S. Goriely}
\affiliation{Institut d'Astronomie et d'Astrophysique,
ULB,  CP 226, B-1050 Brussels, Belgium}

\title{On microscopic theory of radiative nuclear reaction characteristics}


\begin{abstract}
  A survey of some results in the modern microscopic theory of properties of  nuclear reactions with gamma-rays is given. First of all, we discuss the impact of phonon coupling (PC) on the photon strength function (PSF) because it represents the most natural  physical source of  additional strength found for Sn isotopes in recent  experiments  that could not be explained within the standard HFB+QRPA approach. The self-consistent version of the Extended Theory of Finite Fermi Systems  in the Quasiparticle Time Blocking Approximation, or simply QTBA, is applied. It uses the HFB mean field and includes both the QRPA and PC effects on the basis of the SLy4 Skyrme force. With our microscopic E1 PSFs,  the following properties have been calculated  for many stable and unstable even-even  semi-magic Sn and Ni isotopes as well as for double-magic $^{132}$Sn and $^{208}$Pb using the reaction codes EMPIRE and TALYS with several nuclear level density (NLD)  models: 1) the neutron capture cross sections, 2) the corresponding neutron capture gamma spectra, 3) the average radiative widths of neutron resonances. In all the  properties considered, the PC contribution turned out to be significant, as compared with the standard QRPA one,  and necessary to explain the available experimental data. The results with the phenomenological so-called generalized superfluid NLD model turned out to be worse, on the whole, than those obtained with the microscopic HFB+combinatorial NLD model. The very topical question about the M1 resonance contribution to PSFs is also discussed.

Finally, we also discuss  the modern microscopic  NLD models   based on the self-consistent  HFB method and show their relevance to explain experimental data as compared with the  phenomenological models.
The use of these self-consistent microscopic approaches is of particular relevance for nuclear astrophysics, but also for the study of double-magic nuclei. 
\end{abstract}
\maketitle


%
\section{Introduction}
\label{intro}

In order to calculate characteristics of nuclear reactions
with gamma-rays, information on the photon strength function (PSF) and nuclear level
density (NLD) models is necessary. Traditionally, these quantities have
been modeled phenomenologically and the corresponding parameters adjusted on stable nuclei.
Such  information is needed to calculate all characteristics 
of nuclear reactions with gamma-rays, in particular,
the radiative neutron capture cross sections of particular interest in astrophysical \cite{gor04} and nuclear engineering \cite{muhab} applications.
Commonly, one parametrizes the PSF phenomenologically using, for example, Lorentzian-type models \cite{ripl2,ripl3}.
 The usual definition of PSF contains transitions between excited states. 
For this reason, in order to calculate the PSF, the known Brink hypothesis \cite{brink55} is often used which states that on each excited 
state it is possible to build a giant dipole resonance (at present, any giant resonance) including its low-lying part.
In the low-lying energy region, 
there exists the so-called Pygmy Dipole Resonance (PDR). It exhausts typically about 1-2\% of the Energy Weighted Sum Rule (EWSR) but, nevertheless, it can significantly increase
the radiative neutron capture cross section and affect the nucleosynthesis of neutron-rich nuclei by the r-process \cite{gor04}.                                  
  In neutron-rich nuclei, for example, $^{68}$Ni \cite{wiel} and, probably,  $^{72,74}$Ni, the PDR fraction of the EWSR  is expected to be much larger. 
Note that for nuclei with small neutron separation energy, less than typically 3--4 MeV, 
the PDR properties are changed significantly \cite{gor04}, and therefore,  phenomenological systematics obtained by fitting  characteristics of stable nuclei cannot be applied. 
Through the Brink hypothesis \cite{brink55},  the de-excitation PSF is directly connected 
to the photoabsorption cross section  and, therefore, with the PDR field \cite{ripl2, savran2013, kaev2014}. 
For all these reasons, during the last decade there has been an increasing interest in the investigations of the excitations in the PDR
energy region manifested both in "pure" low-energy nuclear physics \cite{savran2013, paar2007}  and in the nuclear data field \cite{gor04, ripl2, ripl3}.
 
 The experiments in the PDR energy region \cite{toft10,toft11,uts2011,tsoneva} have given  additional information about the PDR and PSF structures. The PSF structures at 8--9 MeV in six Sn isotopes obtained  by the Oslo method \cite{toft10,toft11} could not be explained within  the standard phenomenological approach. In order  to explain the experiment, it was necessary to add "by hand" some additional low-lying strength of about 1--2\% of the EWSR.

Given the importance of PSF both in astrophysics \cite{gor04} and nuclear engineering \cite{muhab}, 
microscopic investigations are required, especially when extrapolations to exotic nuclei are needed.
Mean-field approaches using effective nucleon interactions, such as the Hartree-Fock Bogoliubov method and
the quasi-particle random-phase approximation (HFB+QRPA) \cite{gor04}, allow systematic self-consistent 
studies of isotopic chains, and for this reason have been included in modern nuclear reaction codes like EMPIRE \cite{empire} or TALYS \cite{koning12}. Such an approach is of higher predictive power in comparison with phenomenological models. However, as we discuss below, and as confirmed by recent experiments, the  HFB+QRPA  approach is necessary but not sufficient. To be exact, it should be complemented by the effect describing the interaction of single-particle degrees of freedom with the low-lying collective phonon degrees of freedom, known as the  phonon coupling (PC).

 The results in Ref.~\cite{uts2011} directly confirm the necessity to go beyond the HFB+QRPA method because the PSF structures observed in Ref.~\cite{toft11} could not be explained within the  HFB+QRPA approach.  In particular, the PC effects discussed in Refs.\cite{PhysRep,ave2011,kaev2014} may be at the origin of such an extra strength. Note that the microscopic PSFs which contained  transitions  between the ground  and excited states, i.e. the PSFs values at the energy point near the neutron separation energy,  have already been estimated long ago within the quasiparticle-phonon model approach, which also includes the PC \cite{vg}. 
 
 
   There are also some additional questions  for double-magic nuclei. The problem is that  the phenomenological approaches ''smooth'' the individual characteristics 
of these nuclei or consider them on average. Individual  peculiarities are especially expressive just for double-magic nuclei,  even for stable,  not to mention unstable ones.  For example,  to include the vibrational NLD enhancement    to the well-known generalized superfluid model (GSM) \cite{ripl2,  ripl3},
the experimental values for the energies of the first 2$^+$  and the approximate $50A^{-2/3}$ MeV energies for the first 3$^-$ levels  are usually considered. Such a formula is however not suited for double-magic nuclei,  and both these prescriptions  should  not be used for unstable nuclei.  
Microscopic approaches in nuclear theory account for the specificity of each nucleus through its single-particle and collective (phonon) properties.
Therefore,  it allows  for some irregular changes in comparison with global phenomenological models \cite{belanova} to be seen and checked.  Thus,   for double-magic nuclei   it is especially necessary  to use microscopic approaches for both the PSF and NLD. 

Many different NLD models have been developed for the last decades and in many respect NLD predictions are still not satisfactory, especially in view of their importance in reaction modelling. The shortcomings of analytical NLD formulae in matching experimental data are overcome, as a rule, by empirical parameter adjustments. For this reason, their predictive power away from experimental constraints is questionable in contrast to more microscopic approaches that have been seriously developed for the last decades. The present paper also reviews the last attempt made to improve microscopic NLD models for practical applications.  


In this work we discuss and compare i) the  microscopic  and  phenomenological PSF models, ii) the corresponding radiative nuclear reaction characteristics, namely, the neutron radiative capture cross sections, the corresponding capture $\gamma$-ray spectra and average radiative widths, iii)
phenomenological and microscopic NLD models required for calculations of these  radiative nuclear reaction characteristics.  
 To estimate PSFs,  we use   the self-consistent version of the extended theory of finite fermi systems (ETFFS) \cite{PhysRep} in the quasi-particle time blocking approximation (QTBA) \cite{tselyaev}.  
Our ETFFS (QTBA) method, or simply QTBA,  includes self-consistently the QRPA and PC effects and the single-particle continuum in a discrete form. 
Details of the method are described in Ref. \cite{ave2011}. The method allows us 
to investigate the impact of the PC on nuclear reaction in both stable and unstable  nuclei.  We calculate  the microscopic PSFs in several Sn and Ni  isotopes as well as
(due to their above-mentioned specificities) in the doubly-magic nuclei $^{132}$Sn and $^{208}$Pb and use them in the EMPIRE and TALYS codes to estimate  the  neutron capture cross sections, the corresponding capture gamma-ray spectra and the average radiative widths. Finally, in view of the importance of NLD in reaction theory, a review on the last developments made to determine NLD within the mean field plus combinatorial model is given in Sect.~\ref{sect_nld}.

\section{Method}

The strength function $S(\omega) = dB(E1)/d\omega$ \cite{PhysRep,ave2011}, related to  the PSF $f(E1)$  by  
$f(E1,\omega)[{\rm MeV}^{-3}] = 3.487\cdot10^{-7} S(\omega)[{\rm fm}^2{\rm MeV}^{-1}]$,
is calculated by the QTBA method  \cite{PhysRep,tselyaev} on the basis of the SLy4 Skyrme force \cite{chabanat}.  
The ground state is calculated within the  HFB method using the spherical code HFBRAD~\cite{bennaceur}. 
The residual interaction for the (Q)RPA and  QTBA calculations  is derived as the second derivative of the Skyrme functional. In all our calculations
 we use  a smoothing parameter of 200 keV which effectively accounts for correlations beyond the considered PC.
   Such a choice guarantees a proper description of all three characteristics of giant resonances, including the width \cite{PhysRep}, and also corresponds to the experimental resolution of the Oslo method  
   \cite{toft11}.

\section{PSF's and PDR's  }
\subsection{Sn and Ni isotopes}
\begin{figure}
\includegraphics[width=8cm,clip]{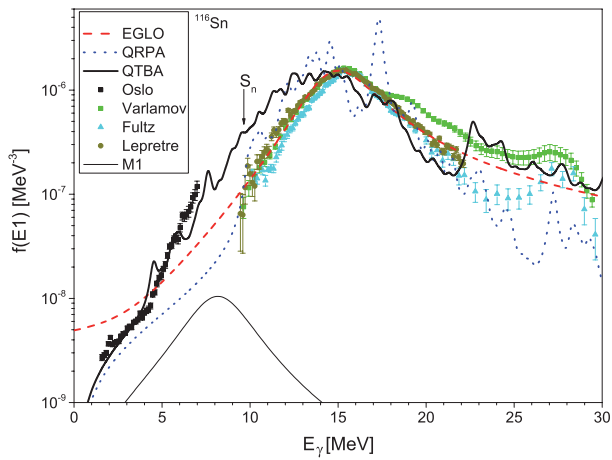}
\caption{ The E1 photon strength functions for $^{116}$Sn. The dashed lines are obtained with the
EGLO model \cite{ripl2}, the dotted line corresponds to the QRPA calculation, and the solid line  to
QTBA calculation. The arrow marks the neutron separation energy $S_n$. The
experimental data are taken from \cite{toft10,Varlamov,Fultz,Lepretre}.}
\label{fig-1}       
\end{figure}

\begin{figure}
\centering
\includegraphics[width=8cm,clip]{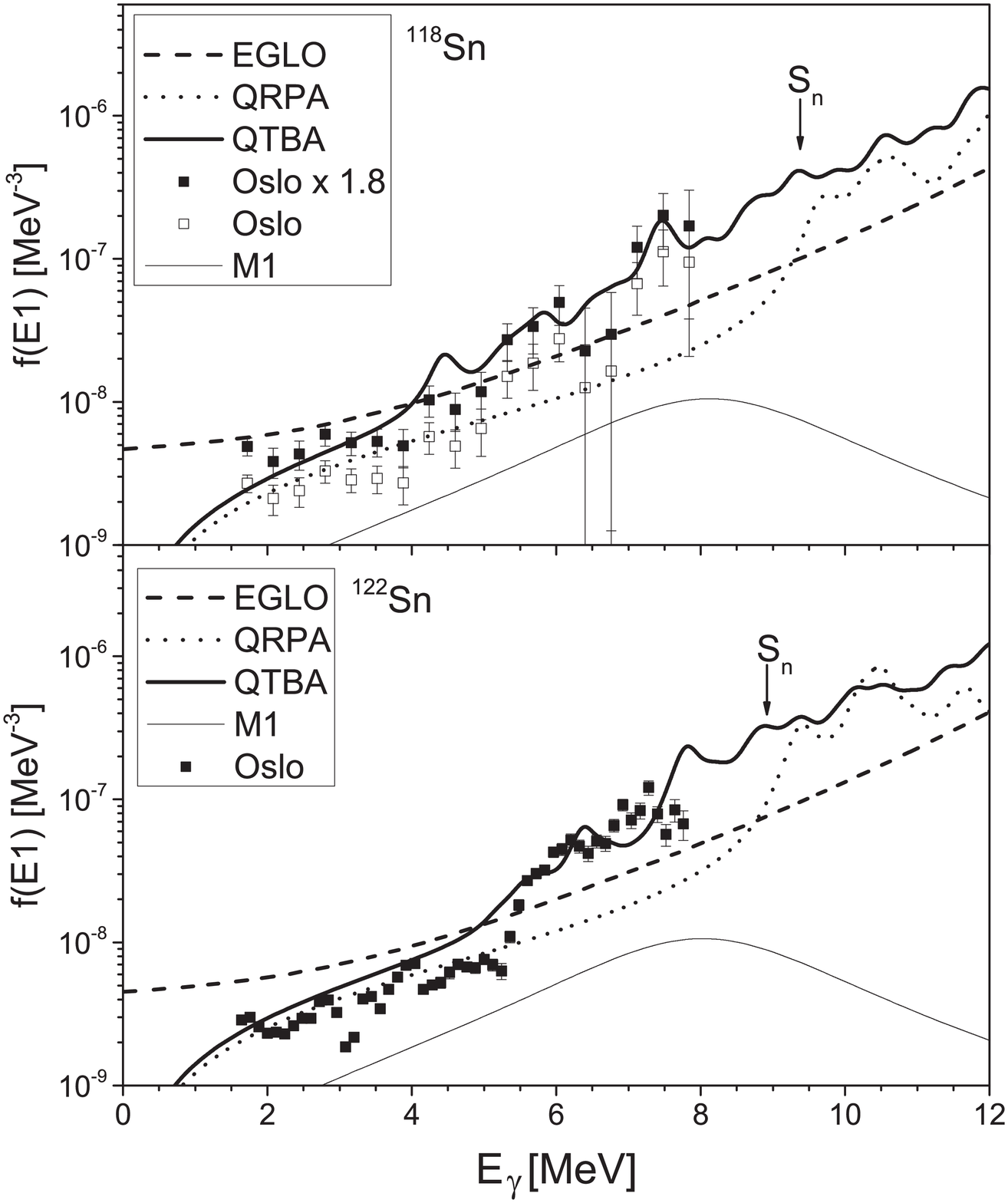}
\caption{Same as Fig. \ref{fig-1}, but for $^{118}$Sn and $^{122}$Sn within the PDR energy region. The
experimental data are taken from \cite{toft10, toft11}.} 

\label{fig-2}       
\end{figure} 

In Figs.~\ref{fig-1} and \ref{fig-2}, the E1 PSFs for three even-even  Sn isotopes are  compared with experimental data obtained with the Oslo method \cite{toft11} for Sn isotopes as well as with the phenomenological Enhanced Generalized Lorentzian (EGLO) model \cite{ripl2}. 
It can be seen that {\it i)} in  contrast  to phenomenological models,  the structure patterns caused by both the QRPA and PC effects are pronounced in both Sn and Ni isotopes. Physically, the PC structures are caused by the  poles at  $\omega = \epsilon_1 -\epsilon_2 - \omega_s $ or $\omega = E_1 + E_2 -\omega_s$, where $ \epsilon_1, E_1, \omega_s $  are single-particle, quasi-particle and phonon energies, respectively. Such a PC effect is seen to become significant above 3~MeV and below typically 9--10~MeV;
{\it ii)} for   $^{118}$Sn and  $^{122}$Sn  isotopes, a reasonable agreement with experiment is obtained  within the QRPA below typically 5 MeV. 
For all three Sn isotopes, at E$>$5~MeV, the inclusion of PC effects is needed to reconcile predictions with experiment \cite{toft11}; {\it iii)} up to the  nucleon separation energy, the phenomenological EGLO description of the experimental data is noticeably worse than the one achieved by the QTBA. 

 \begin{figure}
\includegraphics[width=8cm,clip]{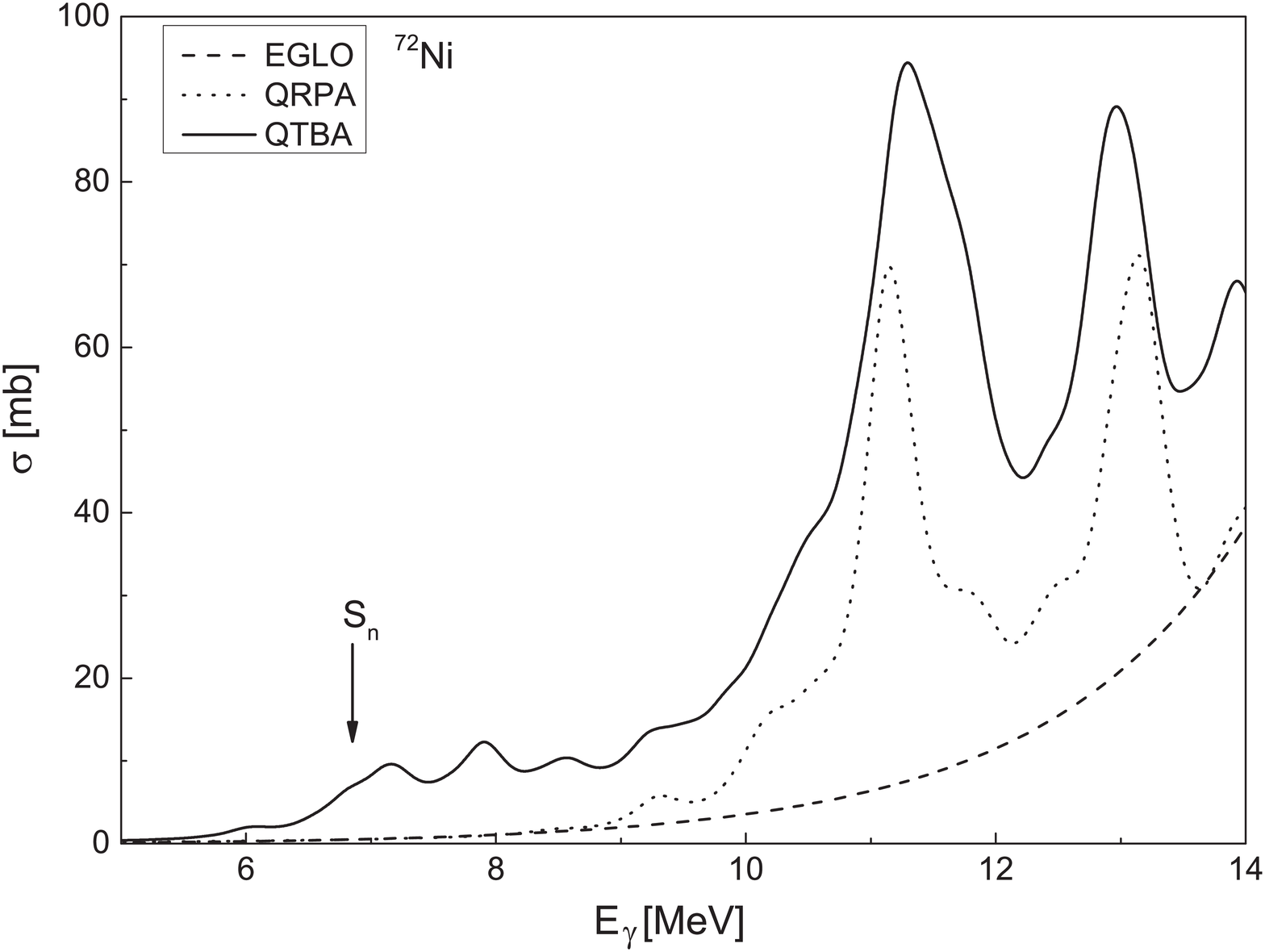}
\caption{The calculated photoabsorption cross section in $^{72}$Ni}
\label{fig-3}   
\end{figure}

In Table \ref{tab-1}, the integral parameters (mean energy $E$ and fraction of EWSR exhausted)  of the PDR are given for  three Ni isotopes, as predicted by both QRPA and QTBA models.   
To compare results  in these three  nuclei,  a 6 MeV energy interval, which corresponds to the one where the PDR was observed in $^{68}$Ni,  is considered. In this interval, the PDR characteristics have been approximated, as usually, with a Lorentz curve by fitting the three moments of the theoretical curves  \cite{PhysRep}.
For $^{68}$Ni, a good agreement is obtained with experimental data of the mean energy $E  \simeq 11$~MeV and about 5\% of the total EWSR  \cite{wiel}. 
A similar calculation was performed for $^{68}$Ni  \cite{lrt2010} using the relativistic QTBA, with  two phonon contributions additionally taking into account.
  For the PDR characteristics in $^{72}$Ni in the (8-14) MeV range, we obtain a mean energy $E=12.4$~MeV,  width $\Gamma = 3.5$~MeV and a large strength of 25.7\% of the EWSR. It should be noted that the main contribution to the $^{72}$Ni PDR is found in the (10-14) MeV interval which exhausts 13.9\% of the EWSR for QRPA and 23.2\% for QTBA. In this interval, two maxima can be observed (Fig~\ref{fig-3}). So the strength in the (10-14) MeV dominates and is globally equivalent to the one in (8-14) MeV.  A significant PC contribution to the PDR strength is found in all isotopes (Table~\ref{tab-1}). 
  
\begin{table}[ht]
\centering
\caption{Integral characteristics of the PDR (mean  energy $E$ in MeV and fraction of the EWSR) in Ni isotopes 
calculated in the (8-14)~MeV interval for $^{58}$Ni, $^{72}$Ni and (7-13)~MeV interval for $^{68}$Ni (see text for details).}
\label{tab-1}
\begin {tabular}{ c c c c c c}
\hline
\hline
\multirow{2}{*}{Nuclei}&\multicolumn{2}{l}{QRPA}&\multicolumn{2}{l}{QTBA}\\
\cline{2-5}
&$E$&\%&$E$&\%\\
\hline
$^{58}$Ni&13.3&6.0&14.0&11.7\\
$^{68}$Ni&11.0&4.9&10.8&8.7\\
$^{72}$Ni&12.4&14.7&12.4&25.7\\
\hline
\hline
\end{tabular}
 \end{table}
 
 \subsection{Double-magic $^{132}$Sn and $^{208}$Pb }



\begin{figure}[t]
\centering
\includegraphics[width=8cm, clip]{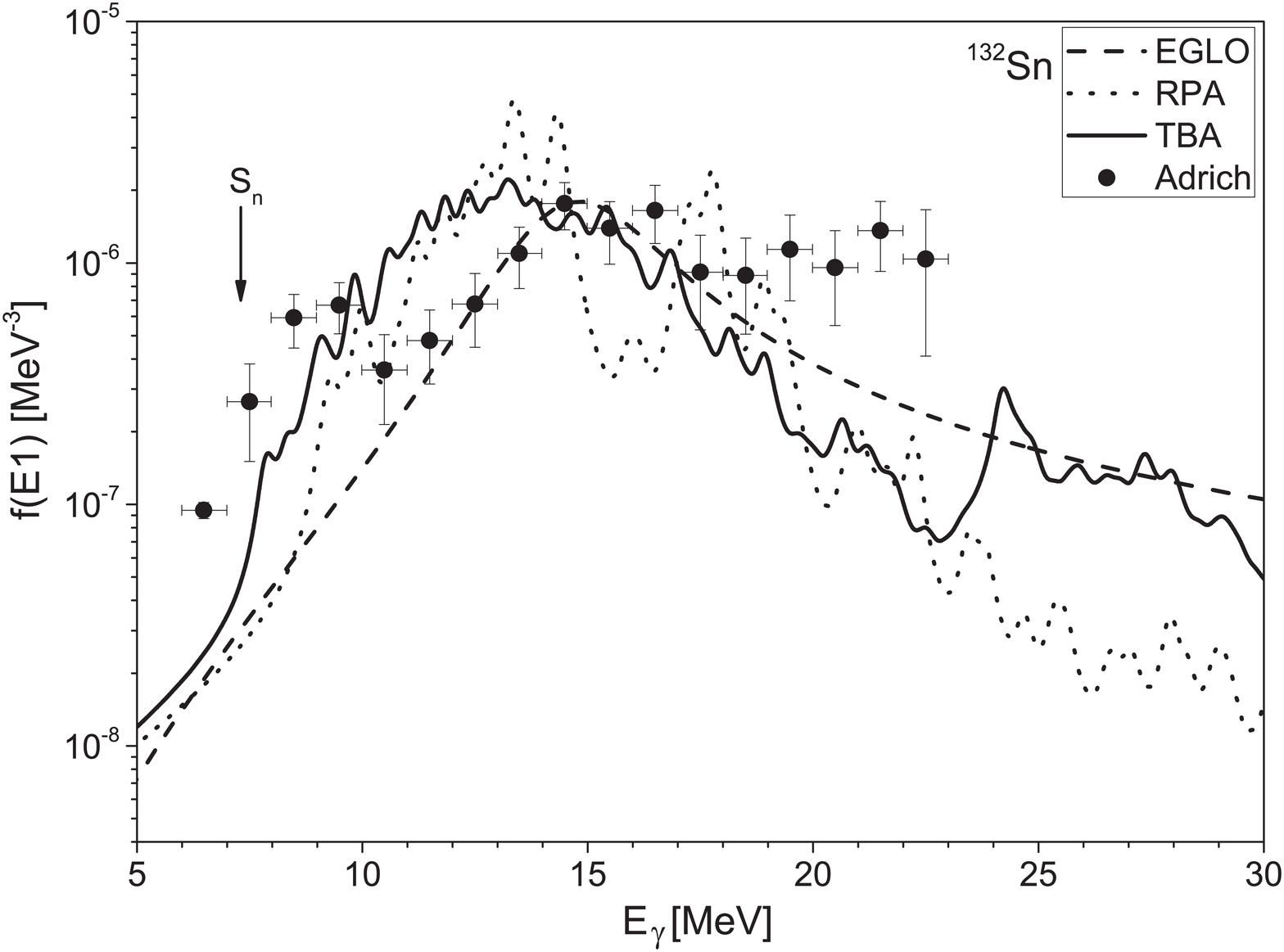}
\caption{The PSF for $^{132}$Sn. Dotted lines correspond to the self-consistent RPA,  solid lines to the TBA  (including PC),  and dashed lines to the EGLO model \cite{ripl2}. Experimental data  \cite{Adrich} were recalculated by us for PSF.}
\label{fig-4}       
\end{figure}

\begin{figure}[t]
\centering
\includegraphics[width=8cm, clip]{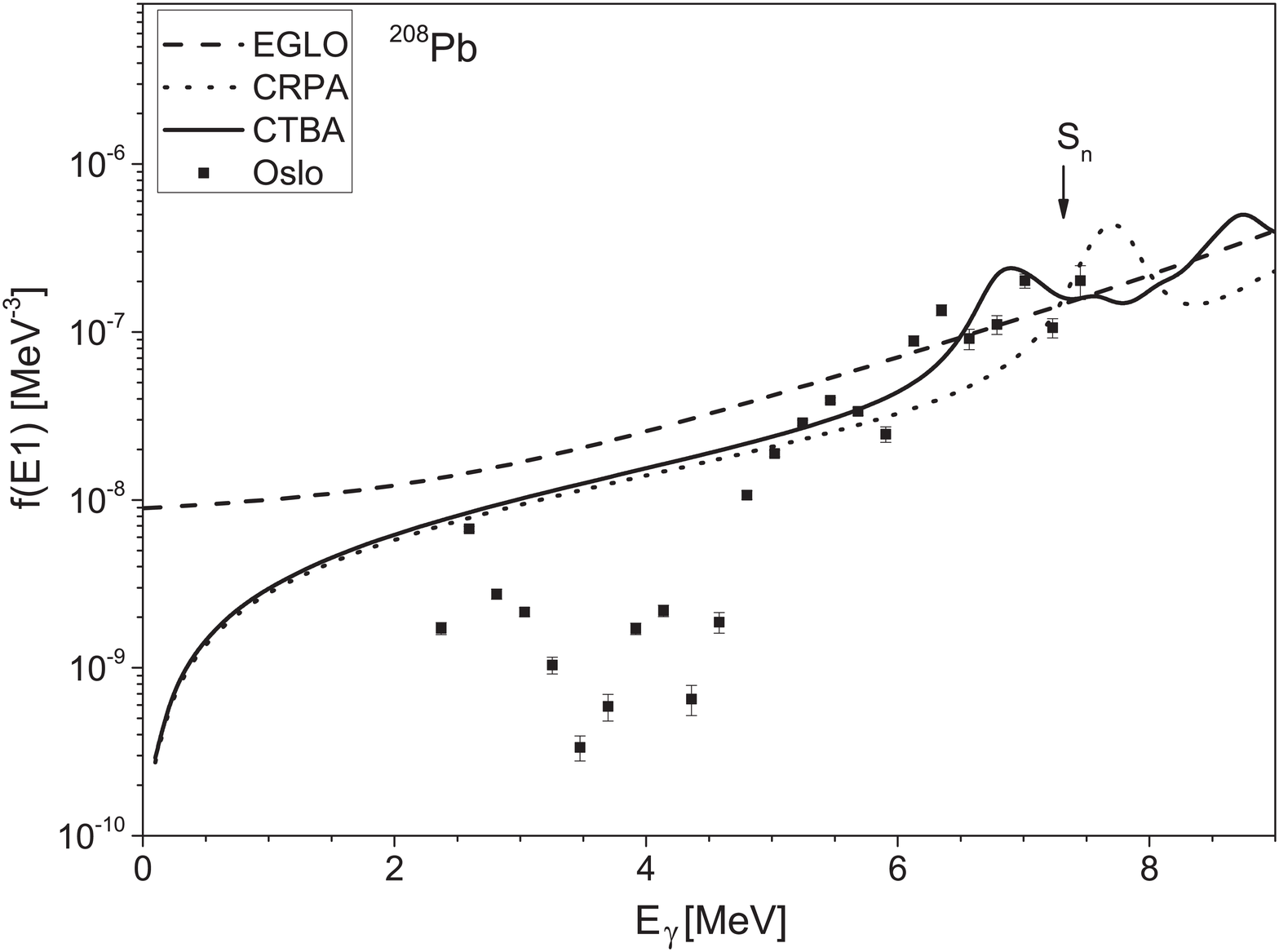}
\caption{ E1 PSF for $^{208}$Pb. Dotted lines correspond to the self-consistent CRPA,  solid lines to the CTBA  (including PC),  and dashed lines to the EGLO model \cite{ripl2}. Experimental data are taken from Ref. \cite{Oslo}.}
\label{fig-5}       
\end{figure}

In Figs.~\ref{fig-4} and \ref{fig-5}, we show the PSFs for $^{132}$Sn and $^{208}$Pb  calculated within  our microscopic TBA  and RPA methods with the SLy4 Skyrme force. These PSFs are  recalculated from the theoretical photoabsorption cross sections taken  from  Ref.~\cite{ave2011}   
  ($^{132}$Sn) and  \cite{TselLut} ($^{208}$Pb). Note that for the first time, the microscopic PSFs are obtained within the
fully self-consistent approach with an exact account for the single-particle continuum (for $^{208}$Pb). The phenomenological EGLO PSFs are also shown.
In Fig.~\ref{fig-4},  the E1 PSF for $^{132}$Sn is compared with experimental data from Ref. \cite{Adrich}. In Fig.~\ref{fig-5},  the E1 PSF for $^{208}$Pb is compared with the experimental data obtained within the Oslo method \cite{Oslo}. As can be seen, in contrast to the phenomenological model EGLO and microscopic continuum RPA,  the CTBA, i.e. RPA + PC, 
can describe some PDR structures in $^{132}$Sn and partly in $^{208}$Pb thanks to the PC effects. For $^{132}$Sn,  we see the well-known structure at about 10 MeV (our approach gives a lower energy), usually referred to as a PDR for the photoabsorption cross section,  as discussed in Refs.~\cite{ave2011, savran2013,  paar2007}.

Let us discuss the results shown in Fig.~\ref{fig-5} for $^{208}$Pb. We see that the CTBA approach describes relatively well experimental data at $E>5$~MeV, at least better than the CRPA model (note that the smoothing parameter 200 keV has been used in the calculations). 
However,  a large  disagreement with experimental data is found at $E<5$~MeV.  As one can see  from Ref.~\cite{rezaeva}, where the transitions between ground and excited states have been measured,  the beginning of the  $1^-$ excitation spectrum is  4.84 MeV,  i.e. there is no $1^-$ transitions between ground and excited states below 4.84 MeV. 
This result is  understandable since in the doubly-magic $^{208}$Pb there is no single-particle  or two-phonon E1 transitions for $E < 5$~MeV.

\begin{figure}
\centering
\includegraphics[width=8cm, clip]{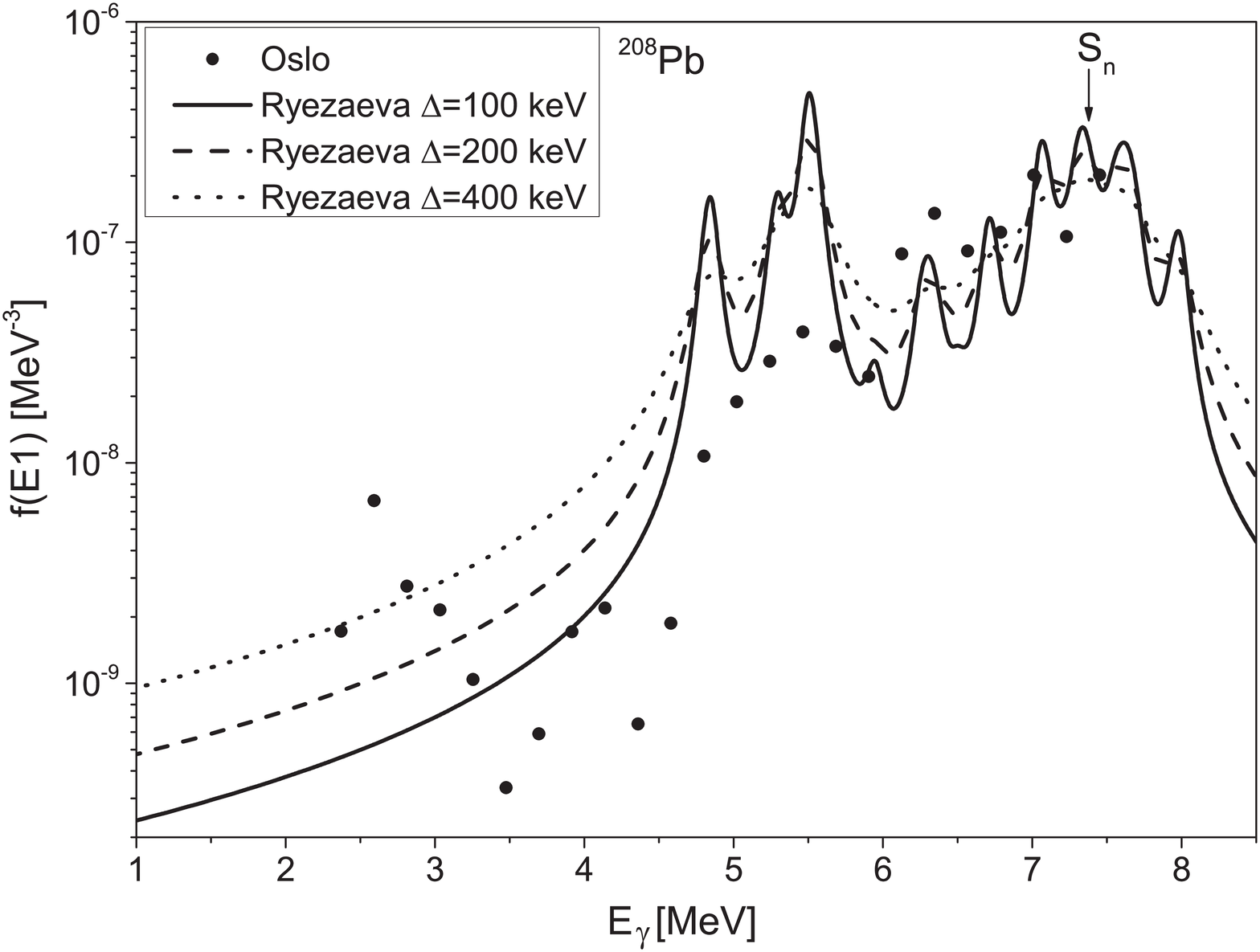}
\caption{ Comparison  of experimental  strengths for $^{208}$Pb: the ($^{3}$He,$^{3}$He$'\gamma)$ reactions method \cite{Oslo} and nuclear resonance fluorescence technique \cite{rezaeva}. The lowest 1$^-$-level of  the data \cite{rezaeva} is 4.84 MeV. It was  smoothed by us with three different values of the smoothing parameter $\Delta$. See text for details.}
\label{fig-6}       
\end{figure}

In order to obtain some  additional information  we  have compared two sets of experimental data for $^{208}$Pb  (see Fig.~\ref{fig-6}): 1) the PSF data from Ref.~\cite{Oslo}  where the transitions between  ground and excited states as well as between excited states contribute and 
2) the data \cite{rezaeva} for the $B(E1)$ values for the transitions between  ground and excited states only.
  It is necessary to compare  both sets of data  with  approximately the same smoothing. So,  taking into account  the 200~keV resolution in the experimental data \cite{Oslo},  we smoothed the data  \cite{rezaeva} with three smoothing parameters 100,  200 and 400 keV according to the following equation
  \begin{equation}
f(E1,\omega)= \frac{8}{27(\hbar c)^3}\sum_{s}B(E1)_s \frac{\Delta}{(\omega-E_s)^2+\Delta^2/4}~.
\end{equation}
  As  expected,  we obtained a rough agreement  between both sets of experimental data  at  $E > 4.84$~MeV. 
  Thus,  one can think that the excitations observed in Ref.~\cite{Oslo} at  $E < 4.84$~MeV are  caused  mainly by transitions between excited states. However,  it is necessary to underline that 
  the mechanisms of the reactions used in Refs.~\cite{Oslo} and \cite{rezaeva} are very different and, what is important here,
 the data from the Oslo $^{208}$Pb  experiment \cite{Oslo} may suffer from a factor of 2 uncertainties due to low level density below the particle separation threshold\footnote{ Private communications with the Oslo group}.

\section{Neutron radiative capture}

\subsection{Semi-magic compound $^{116}$Sn and $^{120}$Sn }

\begin{figure}
\includegraphics[width=8cm,clip]{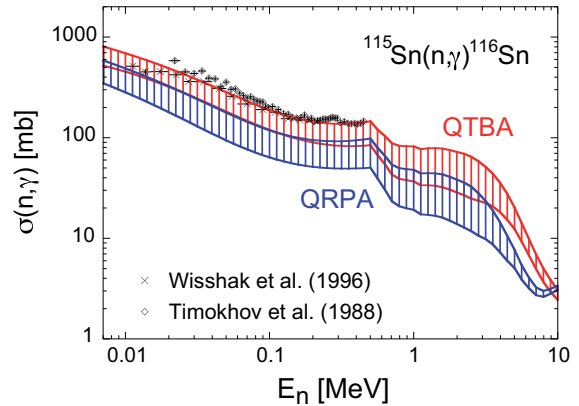}
\caption{$^{115}$Sn(n,$\gamma$)$^{116}$Sn cross section calculated with the  QRPA (blue) and QTBA (red) PSF. The uncertainty bands depict the uncertainties affecting the NLD predictions \cite{ripl2,kon08,gor08,hil12}.
E$_n$ is the neutron energy. Experimental cross sections are taken from Refs.~\cite{wisshak,timokhov}.}
\label{fig-7}   
\end{figure}

\begin{figure}
\includegraphics[width=8cm,clip]{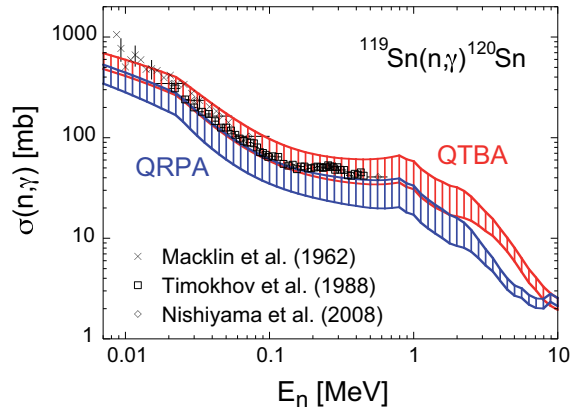}
\caption{ Same as Fig.~\ref{fig-7} for $^{119}$Sn(n,$\gamma$)$^{120}$Sn.  Experimental  cross  section are taken from Refs.~\cite{timokhov,Macklin,Nishi} }
\label{fig-8}  
\end{figure}

In Figs.~\ref{fig-7} and \ref{fig-8}, we  present the radiative neutron capture cross sections  estimated with the Hauser-Feshbach reaction code TALYS \cite{koning12} on the basis of the newly determined gamma-strength function. Similar results are obtained if use is made of the EMPIRE reaction code \cite{empire}. The calculations were  performed with different NLD models, including the back-shifted Fermi gas model \cite{kon08}, the Generalized Superfluid model (GSM) \cite{ripl2} and the HFB plus Combinatorial  model \cite{gor08,hil12} (see Sect.~\ref{sect_nld}). The NLD is constrained by experimental neutron spacings and low-lying states, whenever available \cite{ripl3}.  As seen in Figs.~\ref{fig-7}-\ref{fig-8}, the agreement with experiment is only possible  when the  PC is taken into account \cite{myPRC2015}. QRPA approach clearly underestimates the strength at low energies. This deficiency is often cured by empirically shifting the QRPA strength to lower energies and broadening the distribution \cite{gor02,gor04}.
 
\subsection{Double-magic $^{132}$Sn and $^{208}$Pb}

\begin{figure}
\centering
\includegraphics[width=8cm, clip]{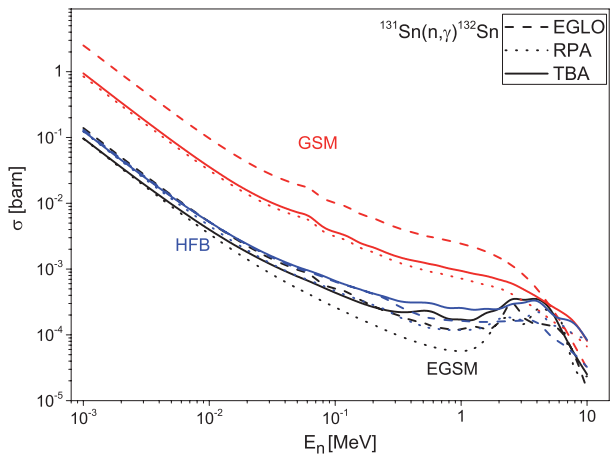}
\caption{ $^{131}$Sn(n,$\gamma$)$^{132}$Sn cross section calculated with the EGLO  (dashed),  RPA  (dotted) and TBA  (solid) PSFs. The red, black and blue curves are obtained with the GSM, EMPIRE-specific and HFB plus combinatorial NLD models, respectively}
\label{fig-9}       
\end{figure}

\begin{figure}
\centering
\includegraphics[width=8cm, clip]{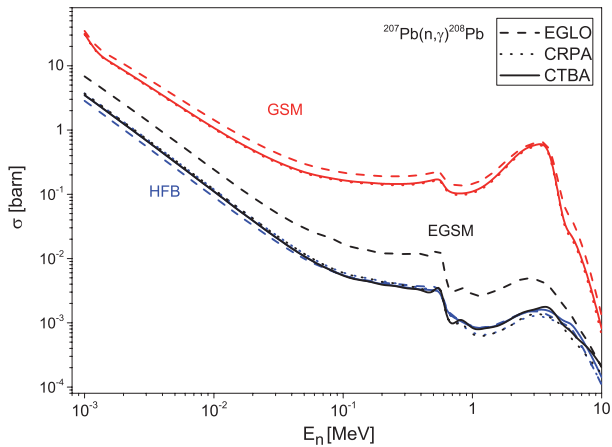}
\caption{ The same as for Fig.\ref{fig-9},  but for the $^{207}$Pb(n,$\gamma$)$^{208}$Pb cross section calculated with the EGLO  (dashed),  CRPA  (dotted) and CTBA  (solid) PSFs.}  
\label{fig-10}       
\end{figure}

In Figs.~\ref{fig-9} and \ref{fig-10} the neutron radiative capture cross sections are shown for the compound $^{132}$Sn and $^{208}$Pb. Note that in Fig.\ref{fig-10}, we do not compare our results with the available $^{207}$Pb(n,$\gamma$)$^{208}$Pb cross section
\cite{Green, Wass} because these low-energy data (two points) are  in the discrete resonance energy region.   We see a  large difference between the results obtained with the GSM and other NLD models  (EMPIRE-specific  and  HFB plus combinatorial),  namely,   the GSM  (n,$\gamma$) cross section   is about one order of magnitude larger for neutrons up to energies of 2 MeV and 10 MeV for the compound $^{132}$Sn and $^{208}$Pb,  respectively. 
There is no noticeable difference between the results obtained with phenomenological EMPIRE-specific and microscopic HFB plus combinatorial NLD models. 
A detailed discussion about these results will be presented somewhere else.
 
\section{Capture gamma-ray spectra}

\begin{figure*}
\centering
\includegraphics[width=15cm,clip]{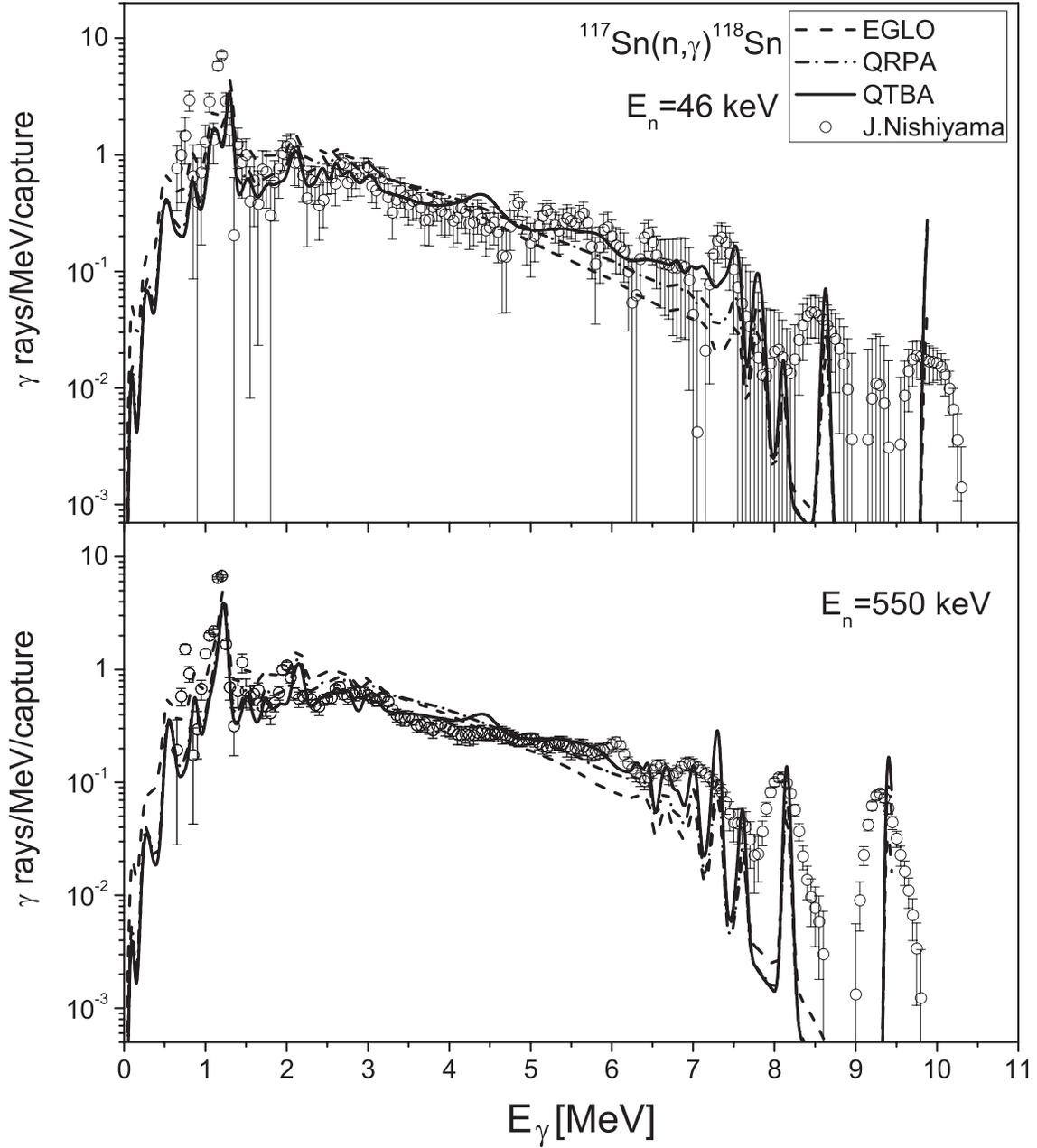}
\caption{Gamma-ray spectra from $^{119}$Sn(n,$\gamma$) for the neutron energies of 52 keV (upper panel)  and 570 keV (lower panel) calculated with EMPIRE and the phenomenological GSM NLD model \cite{ripl2} . Experimental data was taken from \cite{Nishi}. See text for details}
\label{fig-11}       
\end{figure*}
\begin{figure*}
\centering
\includegraphics[width=15cm,clip]{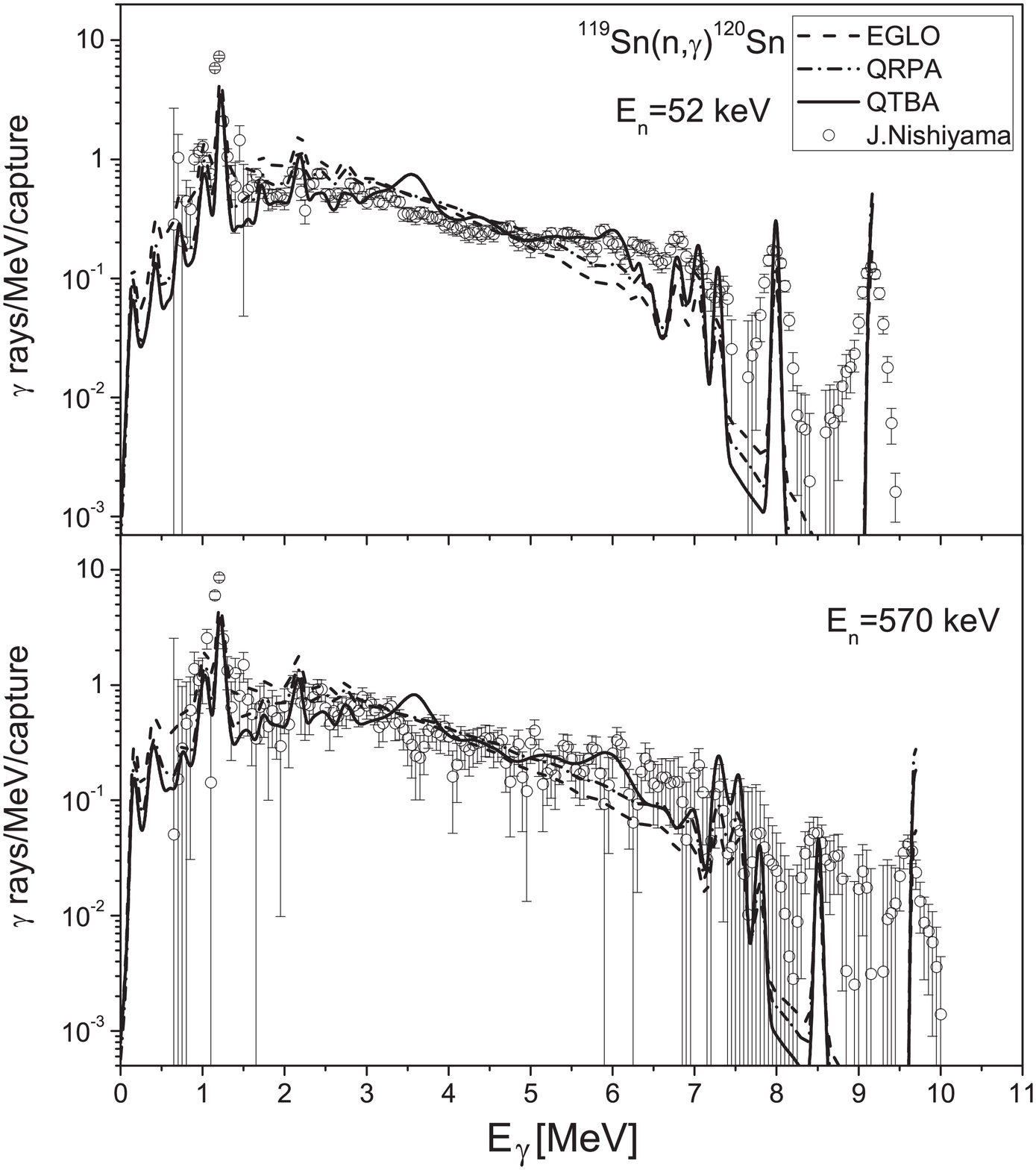}
\caption{Gamma-ray spectra from $^{119}$Sn(n,$\gamma$) for the neutron energy of 52 keV and 570 keV. The microscopic
HFB+combinatorial NLD model \cite{gor08} has been used. Experimental data was taken from \cite{Nishi}.}
\label{fig-12}       
\end{figure*}

In Ref. \cite{Nishi}, neutron capture gamma-ray spectra of $^{117}$Sn and $^{119}$Sn have been measured. With the use of EMPIRE and in order to compare them with theoretical predictions, we divide our results for gamma-ray spectra in mb/MeV units by the capture cross sections at  $\langle E_n\rangle$ = 52 keV and $\langle E_n\rangle$=570 keV for $^{119}$Sn (Fig.~\ref{fig-11}) (The case of $\langle E_n\rangle$=46 keV and $\langle E_n\rangle$ = 550 keV for $^{117}$Sn has been considered in Ref.~\cite{myISINN22_2}). In other words,  we did not consider here the incident neutron energy spectra and took the neutron average energy  of 52 keV given by the authors of Ref.~\cite{Nishi}. The neutron cross sections are calculated with EMPIRE for three theoretical PSF models, EGLO, QRPA and QTBA.  The comparison with  experiment \cite{Nishi} is presented   for two NLD models, namely the GSM  (Fig. \ref{fig-11}) and the  microscopic
HFB plus combinatorial model \cite{gor08} (Fig. \ref{fig-12}).
 One can see from Figs. \ref{fig-11} and \ref{fig-12} that the EGLO and QTBA predictions reproduce experimental data better than QRPA and that the detailed structures are also better described by QTBA than by EGLO.
 
  
As compared with  the phenomenological GSM NLD model,
the agreement with experiment is a little better for the microscopic HFB+combinatorial NLD model.
Our results also show that the PC contribution is noticeable. For all three PSF variants some structures have been  found 
but the QTBA approach describes them a little better. 

\begin{table}[ht]
\centering
\caption{Multiplicities of  capture $\gamma$-rays ($\gamma$-rays/capture) of $^{117,119}$Sn calculated at E$_{\gamma}>$0.6 MeV. For each approach (EGLO, QRPA and QTBA) two NLD models are considered: the phenomenological GSM \cite{ripl2} (first line) and the microscopic HFB plus combinatorial model \cite{gor08} (second line).}
\label{tab-2}
\begin {tabular}{ c c c c c c}
\hline
\hline
Nuclei& E$_n$&EGLO&QRPA&QTBA&Exp.\\
\hline
\multirow{4}{*}{$^{117}$Sn}&\multirow{2}{*}{46 keV}&3.64&3.32&2.99&\multirow{2}{*}{ 3.45 (9)}\\
\cline{3-5}
							&		 &3.86&3.40&3.12&\\
\cline{2-5}
							& \multirow{2}{*}{550 keV}&4.03&3.66&3.39&\multirow{2}{*}{3.80 (20)}\\
\cline{3-5}
							&		 &3.48&4.24&3.73&\\
\hline
\multirow{4}{*}{$^{119}$Sn}&\multirow{2}{*}{52 keV}&3.57&3.23&2.96&\multirow{2}{*}{3.31 (16)}\\
\cline{3-5}
						   &		 &3.74&3.28&3.03&\\
\cline{2-5}
							& \multirow{2}{*}{570 keV}&3.96&3.55&3.26&\multirow{2}{*}{3.66 (19)}\\
\cline{3-5}
							&		 &4.11&3.59&3.33&\\
\hline
\hline
\end{tabular}
 \end{table}

 For completeness and to compare with the available experimental data \cite{Nishi}, the calculation of multiplicity of
observed capture gamma-rays at E$>$0.6 MeV has been performed for two target nuclei $^{117}$Sn and $^{119}$Sn  by integrating the gamma-ray spectra  with respect to the gamma-ray energy (see Table \ref{tab-2}). Two NLD models have been considered, namely the  GSM (first line) and HFB plus combinatorial model (second line). One can see that for the QTBA case  the use of HFB+combinatorial model increases the multiplicity values in the right direction but not significantly. As compared with the QRPA case, the inclusion of the PC decreases the multiplicities. Probably, the account for  the energy spectra of the incident neutrons can improve the agreement with  experiment. 

\begin{figure}
\centering
\includegraphics[width=8cm,clip]{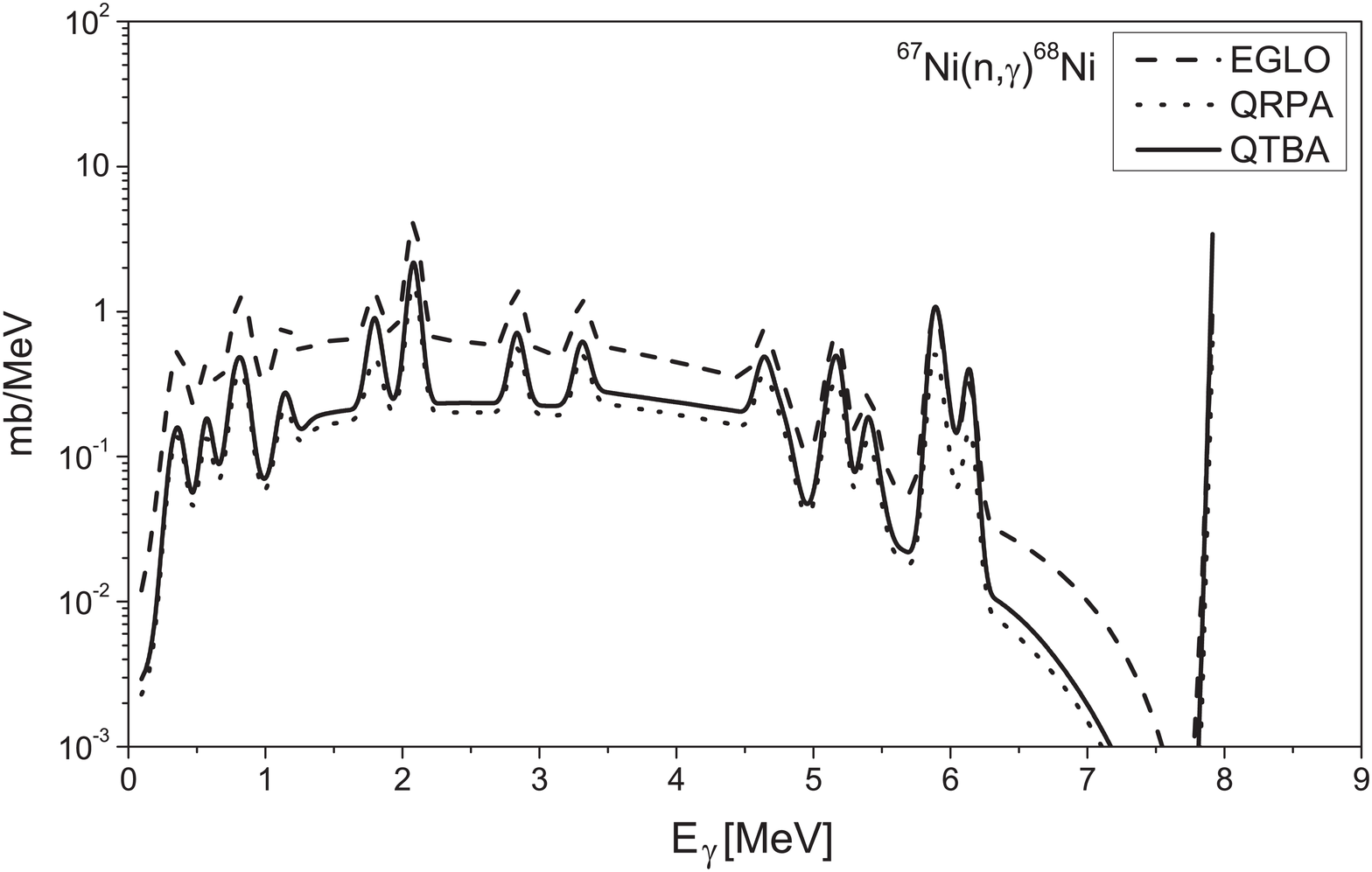}
\caption{Gamma-ray spectra from  $^{68}$Ni(n,$\gamma$) at  the neutron energy of 100 keV. The microscopic
HFB+combinatorial NLD model \cite{gor08} has been used.}
\label{fig-13}       
\end{figure}

We have also performed the capture gamma-ray spectre calculations for the unstable $^{68}$Ni at the neutron energy of 100 keV (see Fig.~\ref{fig-13}).
 Here one can see a large difference between results with our two microscopic  and the phenomenological EGLO  PSF models; this
 confirms the necessity of using the microscopic approach for unstable nuclei. A similar situation is found if use is made of the GSM NLD model in Ref. \cite{myISINN22_2}.

\section{Average radiative widths}

   \begin{table*}[ht]
\centering
\caption{Average radiative widths $\Gamma_{\gamma}$ (meV) for s-wave neutrons. For each approach (EGLO, QRPA and QTBA) two NLD models are considered: the phenomenological GSM \cite{ripl2} (first line) and the microscopic HFB plus combinatorial model \cite{gor08} (second line). See text for details.}
\label{tab-3}
\begin {tabular}{ c c c c c c c c c c c c c c c c}
\hline
\hline
&$^{110}$Sn&$^{112}$Sn&$^{116}$Sn&$^{118}$Sn&$^{120}$Sn&$^{122}$Sn&$^{124}$Sn&$^{132}$Sn&$^{136}$Sn&$^{58}$Ni&$^{60}$Ni&$^{62}$Ni&$^{68}$Ni&$^{72}$Ni&$^{208}$Pb\\
\hline
\multirow{2}{*}{EGLO}&147.4&105.5&72.9&46.6&55.0&56.6&49.9&398&11.1&1096&474&794&166&134&10.6\\
&207.9&160.3&108.9&106.7&124.3&110.2&128.7&4444&295.0&2017&1882&1841&982.2&86.4&2734\\
\hline
\multirow{2}{*}{QRPA}&45.6&34.4&30.4&22.1&23.8&27.9&22.3&133&11.2&358&594&623&75.4&83.8&4.4\\
                     &71.0&49.7&44.3&40.3&43.0&50.1&68.9&4279&447.8&450.8&1646&490.9&406.4&46.7&2973\\
\hline
\multirow{2}{*}{QTBA}&93.5&65.7&46.8&33.1&34.1&35.8&27.9&148&12.3&1141&971&1370&392&154&4.6\\
                     &119.9&87.0&58.4&58.1&61.5&64.0&84.8&4259&509.2&1264&2800&2117&2330&53.8&2448\\
\hline
\multirow{2}{*}{Exp.} \cite{muhab}&&&&117 (20)&100 (16)&&&&&&\multirow{2}{*}{2200 (700)}&2000 (300)&&&\\
$\qquad$ \cite{ripl2}&&&&80 (20)&&&&&&&&2200 (700)&&&\\
\hline
\multirow{2}{*}{M1}&13.0&9.6&8.9&6.1&6.6&7.3&4.9&40.9&1.3&46.1&32&23.2&36.0&49.6&0.79\\
&29.1&18.1&18.5&13.2&13.4&13.1&15.5&341&87.2&17.0&52&31.8&81.6&27.5&5.25\\
\hline
System.&112&109&107&106&105&104&103&85&73&2650&1900&1300&420&320&3770\\
\hline
\hline
\end{tabular}
 \end{table*}   

To test the low-lying strength predicted within the various existing models, we also consider the average radiative widths of neutron resonances 
$\Gamma_{\gamma}$, known to be a property of importance in the description of the $\gamma$-decay from high-energy nuclear states. This quantity is used in nuclear reaction calculations, in particular, to normalize the PSF around the neutron threshold and is defined by \cite{belanova}:
\begin{equation}
 \Gamma_{\gamma} = \sum_{I=|J-1|}^{J+1}\int_{0}^{S_n}\epsilon^3_{\gamma}f_{E1}
 (\epsilon_{\gamma})\dfrac{\rho(S_n-\epsilon_{\gamma},I)}{\rho(S_n,J)}d\epsilon_{\gamma},
\end{equation}
where $\rho$ is the NLD and $J$ the spin of the initial state in the compound nucleus.
Extended compilation of experimental data  for $\Gamma_{\gamma}$ can be found in Refs.~\cite{ripl2,ripl3,muhab}. 

\subsection{Sn and Ni semi-magic isotopes}

We have calculated the $\Gamma_{\gamma}$ values for 13 Sn and Ni isotopes on the basis of the EMPIRE code \cite{empire} for the 3 different PSF models, namely EGLO, our SLy4+QRPA and the present QTBA, together with different NLD prescriptions, namely the GSM \cite{ripl2} and the microscopic HFB plus combinatorial model \cite{gor08}, see Table \ref{tab-3}. The predictions are compared  with experimental data~\cite{muhab}, whenever available, and with existing systematics \cite{ripl2,ripl3}.  
As seen in Table ~\ref{tab-3}, the  PC effect in stable nuclei significantly increases  the QRPA contribution and improves the agreement with the  systematics. Except for $^{122}$Sn and  $^{124}$Sn, where the increase is limited, the PC leads to an enhancement of  about  50 to 200\%. 

 Our  $\Gamma_{\gamma}$  results for $^{118}$Sn, $^{120}$Sn, $^{60}$Ni and $^{62}$Ni, for which  experimental data (not systematics) exists,  are of special  interest.
On the basis of the QTBA strength and the microscopic HFB plus combinatorial NLD \cite{gor08}, we obtain a good  agreement with experiment for $^{60}$Ni, $^{62}$Ni, and reasonable for $^{118}$Sn and $^{120}$Sn.

 Note that on top of the E1 strength, an M1 contribution following the recommendation of Ref.~\cite{ripl3} is included in the calculation of $\Gamma_{\gamma}$. 
The M1 resonance contribution to $\Gamma_{\gamma}$ has been estimated in  Table \ref{tab-3} using the GSM and HFB plus combinatorial NLD models and the standard Lorentzian parametrization \cite{ripl3} with a width $\Gamma = 4$~MeV (note that such a large $\Gamma$ value is open to question, as discussed in Ref.~\cite{kaevkov2006}). Such a contribution is found to be of the order of (10-12)\% of the values in the first line of Table.~\ref{tab-3} for Sn isotopes and 4, 3, 22 and 16\%  for $^{58}$Ni, $^{62}$Ni, $^{68}$Ni and $^{72}$Ni, respectively. In our opinion, the question of the M1 contribution to $\Gamma_{\gamma}$ requires some additional consideration.

The agreement  of the $\Gamma_{\gamma}$ values with experiment is found to deteriorate if use is made of the EGLO or QRPA strengths, but also of the GSM NLD. One can also see  that for stable nuclei, the combinatorial NLD model results  are  in a better agreement with the systematics \cite{ripl2} than those obtained with the GSM model.  As far as the EGLO model is concerned, we see that similar conclusions can be drawn.
 
\subsection{Doubly-magic $^{132}$Sn and $^{208}$Pb}
Unfortunately,  the experimental data are very scarce for the doubly-magic nuclei $^{132}$Sn and $^{208}$Pb. 
However,  as shown in Table \ref{tab-3}, we find, for $^{208}$Pb, 
 a reasonable  agreement between EMPIRE predictions and the $\Gamma_{\gamma}$ systematics \cite{muhab}, at least with the HFB + combinatorial NLD model, not with GSM one (nor with the EMPIRE-specific NLD).  
For the average resonance s-wave level spacings $D_0$, GSM gives 0.0044~keV, EMPIRE-specfic model 32~keV and HFB+Combinatorial model 37.6, the two last to being in agreement with the experimental value of $30\pm 8$~keV. 
The GSM in EMPIRE clearly produces unreasonably small value for $D_0$. EMPIRE also gives  a very small contribution of the M1 resonance to the $\Gamma_{\gamma}$ values for $^{208}$Pb both with the GSM  and HFB+combinatorial NLD models, namely 0.02\% and 0.1\% of the systematics value 
of  $\Gamma_{\gamma}$ = 3770 meV \cite{muhab}, respectively (Table \ref{tab-3}).

\section{NLD models}
\label{sect_nld}

Since Bethe's pioneering work \cite{be36}, many studies have
been devoted to the evaluation of the NLD. The so-called
partition function method is by far the most widely used technique for
calculating level densities, particularly in view of its ability to provide
simple analytical formulae. In its simplest form, the NLD is
evaluated for a gas of non-interacting
fermions confined to the nuclear volume and having equally spaced energy
levels. Such a model corresponds to the zeroth order approximation of a Fermi gas
model and leads to very simple analytical, though unreliable, expressions for the
NLD.
 
In an attempt to reproduce the experimental data,
various phenomenological modifications to the original analytical formulation of
Bethe have been suggested, in particular to allow for shell, pairing and collective 
effects. This led first to the constant temperature formula, then to
the shifted Fermi gas model, and later to the popular back-shifted Fermi gas (BSFG)
model~\cite{gi65,hu72}.  To describe the shell effect, other empirical or
phenomenological improvements to the Fermi formula have been attempted by introducing an
energy dependence to the shell energy shift \cite{ig75,ka78}. To account for a more
realistic pairing correction, analytical expressions have been derived from the BCS
formulation, but at the expense of numerous approximations \cite{go96}. This includes the
GSM approach \cite{ig98} or the semi-classical approximation \cite{go96}.
In such an analytical approach, collective rotational and/or vibrational effects are often not treated explicitly.

Some attempts to include all the shell, pairing and deformation effects
analytically in the NLD formula are found in Refs.~\cite{go96,ig98,ko08}.
 However, drastic approximations are usually made in deriving such analytical NLD formulae
and their shortcomings in matching experimental data are overcome by empirical
parameter adjustments. For practical applications, the available experimental information
can not be omitted and global formulae are tuned by a local determination of the free
parameters for each nucleus. To do so, the NLD formula is renormalized on existing
experimental data (mainly low-lying levels and s-wave neutron resonance spacings). So
far, only simple analytical formulae of the BSFG type have been renormalized and used in 
practical applications \cite{ig98,ko08,di73}.

Several of the approximations used to obtain the NLD expressions in an analytical form
can be avoided by quantitatively taking into account the discrete structure of the
single-particle spectra associated with realistic effective potentials. This approach
leads to the so-called microscopic statistical model (e.g.\cite{de68,mo72,ar81,dem01}).
The NLD computed with this technique is the exact
result that the analytical approximation tries to reproduce. This approach has the
advantage of treating in a natural way shell, pairing and deformation effects on all
the thermodynamic quantities.  However, the statistical model is not free from
uncertainties. In particular  some inherent problems related to the choice of the
single-particle configuration and pairing force remain. The treatment of the collective
enhancement factor as well as the deformation effects at increasing energies and 
for transient nuclei are usually treated in a phenomenological way.
The global microscopic NLD prescription within the statistical approach based on the Hartree-Fock-BCS (HFBCS) 
 ground state properties~\cite{dem01} has proven the capacity of microscopic models to compete with 
 phenomenological models in the reproduction of experimental data and consequently to be adopted for practical 
 applications. However, this statistical approach still presents the drawback of not describing the parity dependence 
 of the NLD, nor the discrete (i.e non-statistical) nature of the excited spectrum at low energies.
 
Other various microscopic models have been proposed, including, for example, the combinatorial model
\cite{hi68,hi97}, the spectral distribution approach \cite{fr71}, Monte-Carlo \cite{ce94} and quantum Monte-Carlo models,  
including correlations beyond the mean-field approximation \cite{alhass97,alhass99,ozen13}. 
Among those,   the combinatorial approach has been the most developed so far and it was demonstrated that such an approach can clearly 
 compete with the statistical approach in the global reproduction of experimental data \cite{gor08,hi01,hi06,hi12}. 
 One of the advantages of this approach is to provide not only the energy, spin and parity dependence of the NLD,
 but also the partial particle-hole level density that cannot be extracted in any satisfactory way from the 
 statistical approaches. At low energies, the combinatorial predictions also provide the non-statistical limit 
 where by definition the statistical approach cannot be applied.

The combinatorial  method consists in using the single-particle level schemes obtained from constrained axially symmetric 
HFB method to construct incoherent particle-hole (ph) state densities as functions of 
 the excitation energy, the spin projection (on the intrinsic symmetry axis of the nucleus) and the parity. Once
 these incoherent ph state densities are determined, collective effects have to be included. In Ref.~\cite{hi06}, the
 choice was made to describe the vibrational effects by multiplying the total level densities by a
 phenomenological enhancement factor  \cite{ig85,ig98} once rotational bands had been constructed.
 The resulting NLD were found to reproduce very well the available experimental data
 (i.e. both the cumulated low energy discrete level histograms and the s- and p-wave resonances mean spacings at
 the neutron binding energy). However, it was clear that the phenomenological treatment of vibrational effects
 needs to be replaced by a sounder treatment. Indeed, rotational bands built on purely vibrational band-heads are
 well established and can clearly not be described if the vibrational enhancement occur once the rotational bands
 are constructed as done in Ref.~\cite{hi06}. Feedback from fission cross section calculations also suggested
 this lack of vibrational states at low energies~\cite{sin07}.
 
  To improve the reliability of the microscopic
 prediction of NLD, the vibrational enhancement factor was later included in the combinatorial approach explicitly
 by allowing for phonon excitations using the boson partition function of Ref.~\cite{hi01} which includes
 quadrupole, octupole as well as hexadecapole vibrational modes. Whereas single-particle levels are
 theoretically obtained for any nucleus, phonon's energies are taken from experimental information when available
 or from analytical expressions~\cite{gor08}. Once the vibrational and incoherent ph state densities are computed,
 they are folded to deduce the total state and the level densities are then deduced exactly like in Ref.~\cite{hi06}.
 To account for the damping of vibrational effects at increasing energies, the folding was restricted to the ph
 configurations having a total exciton number (i.e. the sum of the number of proton and neutron particles and
 proton and neutron holes) $N_{ph} \leq 4$. This restriction stems from the fact that a vibrational state results
 from a coherent excitation of particles and holes, and that this coherence vanishes with increasing number of
 ph involved in the description. Therefore, if one deals with a ph configuration having a large exciton number,
 one should not simultaneously account for vibrational states which are clearly already included as incoherent
 excitations. 
 
 The combinatorial model was further improved \cite{hi12} for deformed nuclei by taking into account 
 the transition to sphericity  coherently on the basis of a temperature-dependent Hartree-Fock calculation. 
 This approach provides at each temperature the structure properties needed to build the level densities. 
 The derived NLDs were shown to provide fairly good results when compared with experimental data \cite{hi12}.

\begin{figure}
\includegraphics[width=9cm,clip]{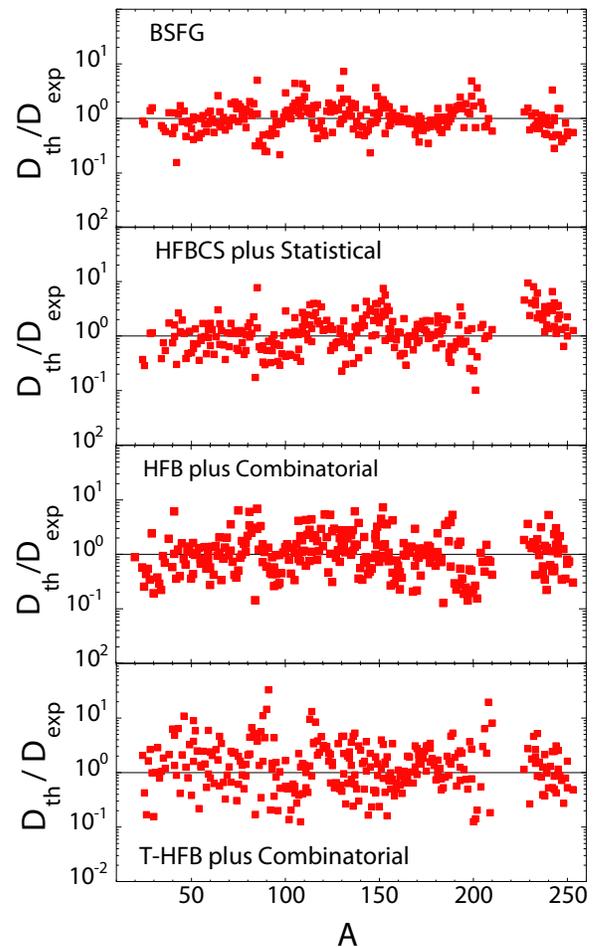}
\caption{ Upper panel: Ratio of BSFG \cite{go02} ($D_{th}$) to the experimental ($D_{exp}$) 
s-wave neutron resonance spacings compiled in Refs.~\cite{ripl2,ripl3}. Middle upper panel: Same with the HFBCS plus Statistical approach of 
Ref.~\cite{dem01}. Middle lower panel: same with the 
HFB plus combinatorial model of Ref.~\cite{gor08}. Lower panel: same with the 
T-dependent HFB plus combinatorial model of Ref.~\cite{hi12}.}
\label{fig-14}  
\end{figure}

As an illustration of the different predictions, we compare in Fig.~\ref{fig-14} the experimental s-wave spacings \cite{ripl2,ripl3}
with those obtained with the BSFG NLD \cite{go02}, the microscopic statistical HFBCS model \cite{dem01} and the HFB plus combinatorial model \cite{gor08}.
Rather similar accuracies (with an average deviation corresponding to a factor of 2) are obtained over the whole nuclear chart where data is available. As shown in Refs.~\cite{hi06,hi12}, the HFB plus combinatorial model also gives satisfactory extrapolations to low energies, 
in particular with respect to the measured cumulative number of levels and the total level density extracted 
by the Oslo group from the analysis of particle-$\gamma$ coincidence in the ($^3$He,$\alpha\gamma$) 
and ($^3$He,$^3$He'$\gamma$) reactions \cite{Oslo}.   

The corresponding NLD tables \cite{gor08} are available to the scientific community at the website {\it http://www-astro.ulb.ac.be}. 
The tables include  the
 spin- and parity-dependent NLD for more than 8500 nuclei ranging from $Z=8$ to $Z=110$  for a large energy and spin grid  ($U=0$ to 200 MeV and the lowest 30 spins). No simple analytical fit to the tabulated NLD is given to avoid losing the specific microscopic characteristics of the model. It should be stressed that the combinatorial NLD cannot be approximated by
 a simple BSFG-type formula, except at very-high energies (above roughly 100~MeV), where the shell, pairing and deformation effects disappear.

One major advantage of the combinatorial method is its non-statistical feature which enables to obtain realistic 
parity and spin distributions, especially at low energies where the statistical limit fails. Another advantage lies in 
the parity dependence which can hardly be estimated within the statistical approach. It should be noted that the 
non-equipartition of parities can have a non-negligible impact, in particular on radiative capture cross sections~\cite{hi06}. 
Concerning the spin distribution, the combinatorial approach provides with NLD that may significantly deviate from 
the usually adopted Wigner law, in particular at low energy since the number of levels is not high enough to reach 
a Gaussian distribution. Such deviations can play a key role when considering the decay to spin isomers at 
low energies~\cite{gok06}, since high-spin populations are usually underestimated within the statistical approach, 
leading to an underestimate of the decaying probability to high spin levels. 
Yet, the combinatorial method can still be improved to better account for collective effects, especially at 
increasing excitation energies. A first attempt has been made to account for the variations of nuclear structure 
properties with increasing excitation energy thanks to a temperature-dependent HFB calculation \cite{hi12}. 
The damping of the vibrational enhancement can also be inspired from the results obtained 
from more microscopic approaches, such as the shell-model Monte Carlo method \cite{ozen13}.

\section{Conclusion}

 The PSFs and PDR  in stable and unstable Ni and Sn isotopes as well as in doubly-magic $^{132}$Sn and $^{208}$Pb 
 have been  calculated within the microscopic self-consistent version of the extended theory of
finite fermi systems which, in addition to the standard QRPA 
approach, includes the PC effects. The Skyrme force SLy4 was used. An update on modern microscopic models of NLD based on the self-consistent  HFB  plus combinatorial  method was also given.

The microscopically obtained  PSFs were compared with the 
available experimental data on E1 strength as well as 
 on nuclear reaction properties using the reaction codes EMPIRE and TALYS. Average radiative
widths, radiative neutron capture cross sections and capture gamma-ray spectra have been calculated taking the PC into account 
 and using various  NLD  models to assess the related uncertainties. 

 The essential and new ingredient of our microscopic QTBA approach is the inclusion of the PC effect.
First, in contrast with phenomenological PSFs, which have no structures, it is
required for the explanation of detailed and observed structures in the PSFs in the energy region below the nucleon separation energy. Second, the quantitative PC contribution 
 significantly improves the agreement with existing experimental data, both for the PSFs, as shown for some Sn isotopes (Figs.\ref{fig-1} and \ref{fig-2}), 
 and for average radiative widths $\Gamma_\gamma$ (Table \ref{tab-3}),
which are increased by about 50-200\% by the PC. 
Here, the  $\Gamma_\gamma$ values are increased always toward better agreement with  experiment or systematics, and
 the PC inclusion 
 results in a reasonable  description of the available experiment.  
We have also shown that the agreement with experimental  $^{115}$Sn(n,$\gamma$)$^{116}$Sn and  $^{119}$Sn(n,$\gamma$)$^{120}$Sn cross sections is only possible  when the  PC is taken into account (Figs. \ref{fig-7} and \ref{fig-8}).

Many observables related to the PSF remain however also sensitive to the adopted NLD model. The HFB plus combinatorial NLD model  can clearly compete with the statistical approach in the global reproduction of experimental  data,  not to mention its higher predictive power.  It provides energy, spin and parity dependence of the NLD  and at low energies describes the non-statistical limit.  We have shown that predictions can be improved with such a microscopic approach, in particular with respect to the phenomenological GSM.

The main conclusion of the work is that  microscopic approaches are necessary to calculate NLD as well as radiative nuclear reaction characteristics and can now replace more phenomenological models, not only for the description and interpretation of experimental data, but also for applications, such as nuclear astrophysics or nuclear engineering.
 
\vspace{0.5cm}

\textbf{ ACKNOWLEDGEMENTS}

The work has been partly supported within  the   cooperation between INPE NRNU MIPHI  and Institute f\"ur Kernphysik FZ J\"ulich.
S.K. acknowledges useful discussions with Drs. V. Furman,
 T.Katabuchi, V.Pronyaev, A.Voinov.  
The authors acknowledge discussions of the $^{208}$Pb results with Therese Renstrom and Profs. S. Siem, M. Guttormsen and A.C. Larsen from the Oslo group. SG is FNRS Research Associate.

\end{document}